\documentclass[aps,twocolumn, superscriptaddress, english, nofootinbib, 10pt, dvipsnames]{revtex4}

\usepackage{float}

\usepackage{amssymb, amsmath, bm, dcolumn, epsf, graphicx, latexsym, slashed, simplewick}
\usepackage[utf8]{inputenc}
\usepackage{subfigure}
\usepackage{comment}
\usepackage{graphicx}
\usepackage{hyperref}
\usepackage{enumerate}

\newcommand{\ba}{\begin{eqnarray}}
\newcommand{\ea}{\end{eqnarray}}

\begin{document}

\title{Constraints on Pre-Recombination Early Dark Energy from SPT-3G Public Data}

\date{}

\author{Adrien La Posta}
\affiliation{Universit\'{e} Paris-Saclay, CNRS/IN2P3, IJCLab, 91405 Orsay, France}

\author{Thibaut Louis}
\affiliation{Universit\'{e} Paris-Saclay, CNRS/IN2P3, IJCLab, 91405 Orsay, France}

\author{Xavier Garrido}
\affiliation{Universit\'{e} Paris-Saclay, CNRS/IN2P3, IJCLab, 91405 Orsay, France}

\author{J.~Colin Hill}
\affiliation{Department of Physics, Columbia University, New York, NY 10027, USA}
\affiliation{Center for Computational Astrophysics, Flatiron Institute, New York, NY 10010, USA}

\begin{abstract}
Early dark energy (EDE) is a proposed solution to the Hubble tension in which a new cosmological field accelerates cosmic expansion prior to recombination and reduces the physical size of the sound horizon. In previous work, a slight preference for a non-zero EDE contribution was found in the latest Atacama Cosmology Telescope data (ACT DR4), while the Planck satellite legacy data alone do not show evidence for it. In this work, we use the most recent public data from the South Pole Telescope (SPT-3G) to constrain the parameters of the EDE scenario. We find that at the current precision level of SPT-3G, an EDE contribution to the total energy density of the universe prior to recombination of $\sim 10\%$ can not be ruled out, but that the data are also consistent with no EDE. The combination of ACT DR4 and SPT-3G with the Planck large-scale temperature anisotropy measurement shows a hint ($2.6\sigma$) for non-zero EDE; however, this preference disappears when the full Planck 2018 data set is included. 

\end{abstract}

  \date{\today}
  \maketitle

\section{Introduction}\label{sec:intro}

With the most recent results from the SH0ES collaboration \cite{2021arXiv211204510R}, the tension in the measurements of the Hubble parameter $H_{0}$ from Planck Cosmic Microwave Background (CMB) data \cite{Planck2018:cosmo} and Cepheid-calibrated SNIa distances has reached a statistical significance of $5\sigma$, with $H^{\rm Planck}_{0} = 67.36 \pm 0.54$~km/s/Mpc and $H^{\rm SH0ES}_{0} = 73.04 \pm 1.04$~km/s/Mpc\footnote{Note that accurate quantification of the tension requires use of the full (non-Gaussian) posterior from the SH0ES $H_0$ measurement.}.  It is important to note that the Planck measurement assumes the $\Lambda$CDM cosmological model, while the SH0ES measurement does not. \\

In the meantime, independent measurements by ground-based CMB experiments, including the Atacama Cosmology Telescope (ACT)~\cite{Choi2020} and the South Pole Telescope (SPT-3G)~\cite{dutcher2021measurements}, have confirmed the preference for a lower value of the Hubble constant inferred from the CMB when assuming the standard $\Lambda$CDM model, with $H^{\rm ACT}_{0}=68.4 \pm 1.5$~km/s/Mpc~\cite{Aiola2020,hill2021} 
and $H^{\rm SPT-3G}_{0} =68.8 \pm 1.5$~km/s/Mpc~\cite{dutcher2021measurements}. In addition, a re-analysis of the Planck data using the Pearson correlation coefficient of T and E modes $\mathcal{R}^\mathrm{TE}_{\ell}$, an observable insensitive to multiplicative systematics, has lead to another robust determination of $H^{\rm Planck, \mathcal{R}^{TE}_{\ell}}_{0}= 67.5 \pm 1.3$~km/s/Mpc \cite{Louis2019, Laposta2021}. \\

Other direct low-redshift probes have inferred values of $H_{0}$ that agree with the value predicted in $\Lambda$CDM from CMB data, in particular the latest strong lensing time delay data \cite{Birrer2020} and the tip of the red giant branch-calibrated SNIa \cite{Freedman2019}.  However, the uncertainties associated to these measurements are sufficiently large that they are still compatible with the measurement from SH0ES. \\

This situation has lead to the development of many extensions of the standard $\Lambda$CDM model attempting to resolve the tension by increasing the value of $H_0$ inferred from CMB and large-scale structure data. A large fraction of these models works by reducing the comoving sound horizon at recombination $r_{*} = \theta_{*} D_{M}  $ (see, e.g.,~\cite{Knox2020} for a review). The angular size of the sound horizon $\theta_{*}$  being directly measured by CMB experiments, this reduction leads to a decrease in the comoving angular diameter distance $D_{M}$ and therefore an increase of the $H_{0}$ value inferred from these experiments. \\

Here we focus on a particular extension, Pre-Recombination Early Dark Energy (EDE) \cite{smith2020, poulin2019, 2019PhRvD.100f3542L, hill2020}. In this scenario, a new cosmological (pseudo) scalar field becomes relevant prior to recombination and accelerates cosmic expansion. This acceleration leads to a change of the conformal time at decoupling and reduces the distance traveled by sound waves in the primordial Universe.  \\

Previous studies have already demonstrated that EDE is a viable candidate to solve the Hubble tension~\cite{poulin2019,Agrawal2019}, albeit at the cost of worsening tensions between CMB- and large-scale-structure-based inferences of the amplitude of fluctuations~\cite{hill2020,Ivanov2020,2021JCAP...05..072D}, which has spurred attempts to fully restore concordance by further modifying the EDE scenario (e.g.,~\cite{AdsEDE20,Karwal2021,ye21EDE,McDonough2021,Clark2021}). The presence of the EDE component significantly reduces the $\chi^2$ for the joint fit of Planck data + SH0ES with respect to $\Lambda$CDM \cite{poulin2019}. It was also recently shown that while the Planck data alone do not favor large value of EDE \cite{hill2020}, high-resolution data from ACT has a mild preference for non-zero EDE \cite{hill2021} (this was found also in subsequent analyses~\cite{poulin2021,Moss2021}; see also Ref.~\cite{2020PhRvD.102l3523L} for related work).  \\

 In this work, we constrain the EDE model using the latest data from the South Pole Telescope, SPT-3G \cite{dutcher2021measurements}, another high-resolution CMB experiment located at the South Pole.  Our goal is to provide an independent test of the hints for EDE seen in the ACT DR4 data. \\
 
 The paper is organized as follows. In Section \ref{sec:data} we describe the different data and likelihoods used to constrain the EDE model. In Section \ref{sec:results} we report the different results we obtain for the posteriors of the cosmological parameters of the $\Lambda$CDM + EDE model. We conclude in Section \ref{sec:conclusion}.

\section{Data and Likelihoods}\label{sec:data}

In addition to parameter posteriors from the latest SPT-3G data, we will display parameter constraints including data from ACT, Planck, and large-scale structure surveys. In this section we describe the data and likelihoods used for this work.

\subsection{SPT-3G}\label{subsec:spt}
We use the most recent SPT-3G power spectra from the 2020 data release~\cite{dutcher2021measurements}, which includes only TE and EE spanning multipoles $300<\ell<3000$. The data have been collected during just half of a typical season of observation and with only part of the focal plane operational. The SPT-3G data release includes a \textsc{fortran} likelihood characterizing these spectra; we present here and use throughout a \textsc{python} version of this likelihood\footnote{Made available at \url{https://github.com/xgarrido/spt_likelihoods}}. We have verified that our \textsc{python} implementation leads to the same results as the official SPT-3G constraints published in Ref.~\cite{dutcher2021measurements}. For $\Lambda$CDM, the recovered parameter posteriors agree to better than 0.1$\sigma$. The likelihood includes modelling of polarized Galactic dust both for TE and EE and Poisson-distributed point sources in EE for the three frequency channels at 95, 150, and 220~GHz.

\subsection{ACT}\label{subsec:act}
We use the latest ACT data, which includes multifrequency temperature and polarization measurements from the fourth data release, ACT DR4~\cite{Choi2020}. The TT, TE, and EE power spectra have been marginalized over foregrounds and systematic uncertainties and are contained in the publicly available \textsc{pyactlike} likelihood,\footnote{\url{https://github.com/ACTCollaboration/pyactlike}} which also uses data from the ACT MBAC DR2 data set. The TE and EE power spectra cover multipoles $326<\ell<4325$, while the TT power spectrum spans $576<\ell<4325$ ~\cite{Aiola2020}. 
The likelihood depends on only the cosmological parameters (6 for $\Lambda$CDM + 3 more for EDE) and one nuisance parameter, the overall polarization efficiency.

\subsection{Planck}\label{subsec:planck}
As our baseline for Planck we use the multifrequency  TT, TE, and EE power spectra and covariances from the Planck 2018 legacy release (PR3)~\cite{P18_like} included in the publicly available \textsc{plik} high-$\ell$ likelihood. We also use the low-$\ell$ TT likelihood, but we do not include low-$\ell$ EE data from Planck, and instead impose a constraint on the optical depth $\tau$ via a Gaussian prior $\tau = 0.065\pm0.015$ (as in Ref.~\cite{Aiola2020}). Following Ref.~\cite{hill2021}, for some of the runs, we use a subset of Planck data consisting only of the TT power spectrum up to a multipole of $\ell <650$. Unlike smaller angular scales, these multipoles are poorly constrained by ground-based experiments like ACT and SPT due to the high level of atmospheric contamination. This data subset can also be seen as an emulation of the WMAP measurements, which was shown to agree very precisely with Planck on these angular scales~(see Fig.~48 of Ref.~\cite{2016A&A...594A..11P}). For brevity, we will refer to this Planck data subset as “PlanckTT650”.

\subsection{Large-Scale Structure Probes }\label{subsec:LSS}
In addition to the results using only primary CMB data, some of the constraints presented in this paper also include large-scale structure probes, in particular CMB lensing and BAO data. The addition of these data helps in breaking degeneracies between cosmological parameters measured solely from the primary CMB anisotropies. \\

\begin{enumerate}
    \item Planck CMB lensing: gravitational lensing of the CMB has been detected with a high statistical significance ($40\sigma$) using Planck data \cite{Plancklensing2018}, which serves to probe the growth of structure over a broad range of redshifts $z \approx 1-2$. The modes included in the reconstruction lie within the range $ 8 \leq  L \leq 400 $.
    \item Baryon Accoustic Oscillation (BAO) data: galaxy surveys and in particular the measurement of the BAO peak serve as a probe of the cosmic expansion history at low redshift. We use data from the SDSS BOSS DR12 LOWZ and CMASS galaxy samples at $z = 0.38$, 0.51, and 0.61 \cite{BOSSDR12}, from the 6dF galaxy redshift survey  at $z = 0.106$ \cite{6dF}, and from the SDSS DR7 main galaxy sample at $z = 0.15$ \cite{SDSSDR7}.
\end{enumerate}

\section{Results}\label{sec:results}

In this section we discuss the constraints obtained on the EDE model from SPT-3G data and various combinations with other CMB datasets. We sample from the parameter posterior distributions using the Monte Carlo Markov Chain (MCMC) method implemented in \textsc{Cobaya}~\cite{Torrado21, cobaya_soft}. To assess the convergence of the chains, we use the Gelman-Rubin convergence test, using the criterion $R-1<0.03$. We also use the \textsc{Bobyqa} likelihood maximizer~\cite{Cartis2018a, Cartis2018b} from \textsc{Cobaya} to obtain a best-fit value for each sampled parameter. Posterior distributions and 68/95\% confidence intervals are obtained from \textsc{GetDist}~\cite{getdist}. The power spectra are computed using \textsc{CAMB}~\cite{Lewis2000,Howlett2012} for which the EDE model is implemented as \texttt{EarlyQuintessence}\footnote{\url{http://camb.info}}. In Appendix~\ref{app:camb_class}, we show that we are able to reproduce the results for the ACT DR4 + P18TT650 dataset from Ref.~\cite{hill2021}, which were computed using a modified version of the \textsc{CLASS}~\cite{Blas2011} Boltzmann solver, \texttt{CLASS\_EDE}~\cite{hill2020}\footnote{\url{https://github.com/mwt5345/class_ede}}. \\

We use the \textsc{CAMB} default precision settings for our MCMC runs, and we increase the accuracy settings as described in Ref.~\cite{hill2021} for obtaining the best-fit values using the \textsc{Bobyqa} minimizer (see also Ref.~\cite{McCarthy2021} for further discussion related to Boltzmann accuracy settings).\\

The EDE component is implemented as an axion-like scalar field, evolving in the potential $V_n(\phi) = m^2f^2(1-\mathrm{cos}(\phi/f))^n$ with $m$ a mass parameter, 
$f$ the axion decay constant of the field, and $n=3$ for the standard EDE scenario considered in most previous analyses.  Following previous studies of EDE, we use an effective parametrization with $z_c$, the redshift at which the EDE field reaches its maximum contribution to the total energy budget, and $f_\mathrm{EDE}(z_c) = \Omega_\mathrm{EDE}(z_c)/\Omega_\mathrm{tot}(z_c) \equiv f_\mathrm{EDE}$ the fractional energy density of EDE at redshift $z_c$. The last parameter is the scalar field initial value $\theta_i = \phi_i/f$. \\

In this work, we thus explore a parameter space defined by the three extra parameters introduced in the EDE scenario $\{f_\mathrm{EDE}, \mathrm{log}_{10}(z_c), \theta_i\}$ and by the $\Lambda$CDM parameters $\{\theta_\mathrm{MC}, \Omega_bh^2, \Omega_ch^2, \mathrm{ln}(10^{10}A_s), n_s, \tau\}$. We choose to use wide uninformative priors on the cosmological parameters, except for the reionization optical depth, for which we impose a Gaussian prior instead of using large-scale polarization data from Planck. We also use wide flat priors on the EDE parameters, allowing for higher values of the critical redshift $z_c$ with respect to what was assumed in Refs.~\cite{smith2020, hill2021, hill2020, Ivanov2020}. We discuss the effect of this prior in detail at the end of Section~\ref{subsec:sptdata}.  The exact priors used in this paper are shown in Table~\ref{tab:priors}. \\

\begin{table}[htpb!]
\begin{tabular}{|l|c|}
\hline
\textbf{Parameters} & \textbf{Prior}\\ 
\hline\hline
\bm{$f_\mathrm{EDE}$} & $[0.001, 0.5]$\\
\bm{$\mathrm{log}_{10}z_c$} & $[3, 5]$\\
\bm{$\theta_i$} & $[0.1, 3.1]$\\
\hline
\bm{$100\theta_\mathrm{MC}$} & $[1.03, 1.05]$\\
\bm{$\log(10^{10}A_\mathrm{s})$} & $[1.6, 3.9]$\\
\bm{$n_\mathrm{s}$} & $[0.8, 1.2]$\\
\bm{$\Omega_\mathrm{b}h^2$} & $[0.01, 0.03]$\\
\bm{$\Omega_\mathrm{c}h^2$} & $[0.05, 0.3]$\\
\bm{$\tau_\mathrm{reio}$} & $\mathcal{N}(0.065, 0.015)$\\
\hline
\end{tabular}
\caption{\label{tab:priors} Prior distributions for the EDE and $\Lambda$CDM parameters. The reionization optical depth prior is a Gaussian probability density centered at 0.065 with a 0.015 standard deviation.}
\end{table}

In order to quantify the preference in favor of the EDE model with respect to $\Lambda$CDM, we will make use of the Akaike Information Criterion (AIC) \cite{Akaike1998} defined by $\mathrm{AIC} = \chi^2_\mathrm{min} + 2N_\mathrm{p}$ with $N_\mathrm{p}$ the number of sampled parameters in the model. Then, the difference of AIC between the EDE and $\Lambda$CDM constraints, $\Delta\mathrm{AIC} = \Delta\chi^2_\mathrm{min} + 2\Delta N_\mathrm{p}$, contain a penalty proportional to the number of extra parameters introduced by the EDE model $\Delta N_\mathrm{p} = 3$. We also quote a level of preference in units of ``sigmas'' with the associated probability, computed from the fact that the $\Delta \chi^2$ for EDE compared to $\Lambda$CDM follows a $\chi^2$ distribution with $\Delta N_\mathrm{p} = 3$ degrees of freedom.

\subsection{Constraining EDE from SPT-3G data}\label{subsec:sptdata}

\begin{figure*}[!p]
\includegraphics[width=\textwidth]{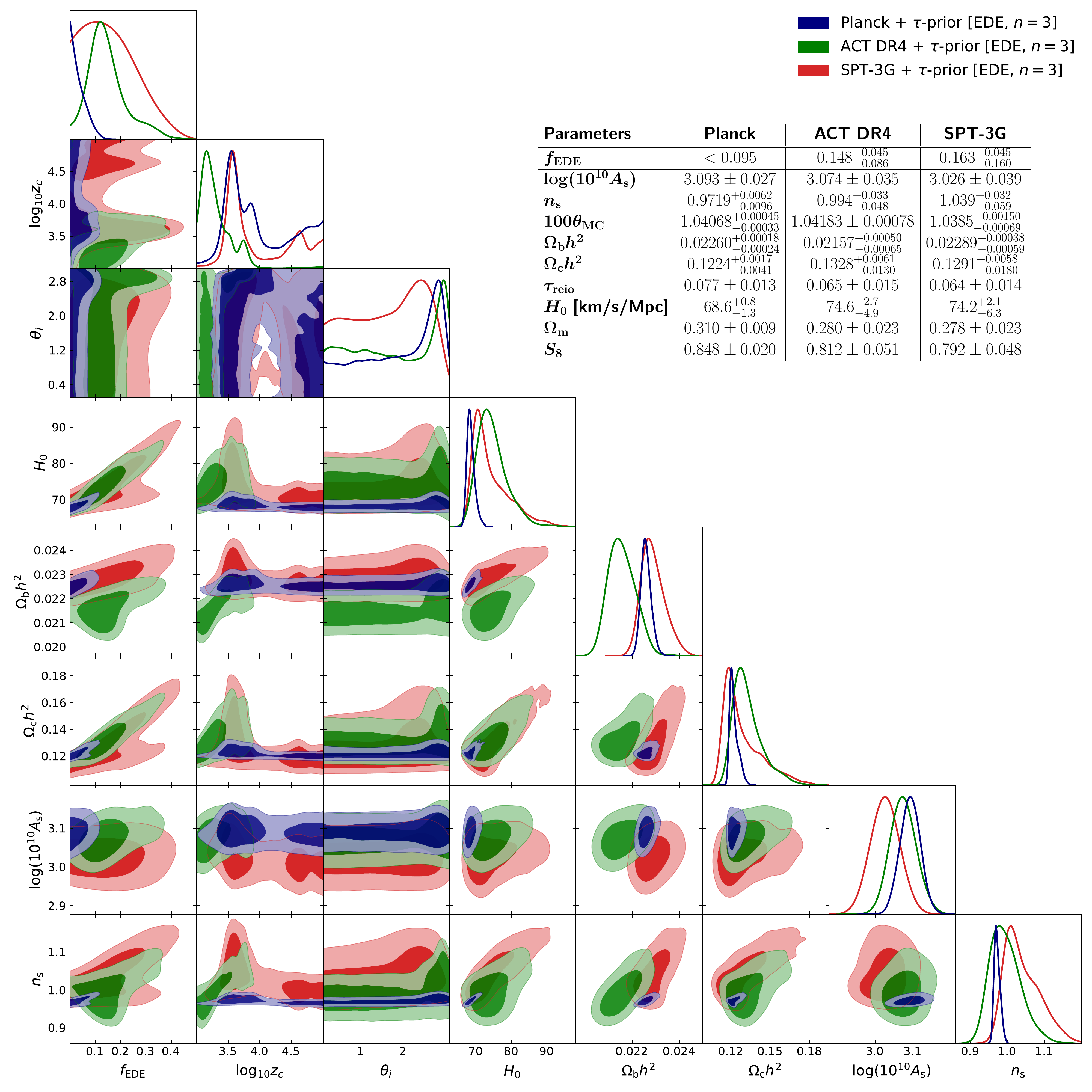}
\caption{Marginalized posterior distributions of EDE ($n=3$) and $\Lambda$CDM parameters derived from Planck 2018 (TT+TE+EE) high-$\ell$ + Planck 2018 low-$\ell$ (TT) data (dark blue); ACT DR4 (TT+TE+EE) data~\cite{Choi2020,Aiola2020} (green); and SPT-3G (TE+EE) latest polarization measurements~\cite{dutcher2021measurements} (red). We impose the same Gaussian prior on the reionization optical depth $\tau$ for all the analyses. We display the 68\% marginalized constraints and the upper/lower limits (95\% C.L.) for the EDE and $\Lambda$CDM in the table associated with the figure.}\label{fig:spt_act_p18}
\end{figure*}

First, we constrain the EDE model using the three most constraining CMB datasets today: Planck 2018, ACT DR4, and SPT-3G. Posterior distributions for the EDE and $\Lambda$CDM parameters are shown in Fig.~\ref{fig:spt_act_p18} and marginalized constraints are available in the associated table. As found in previous studies, Planck data alone do not favor a large fraction of EDE, $f_{\rm EDE} < 0.095 $ ($95\%$~C.L.), while ACT data display a slight preference for non-zero EDE, $ f_{\rm EDE} = 0.148^{+ 0.045}_{-0.086} $. Our new result for SPT-3G, $f_{\rm EDE} = 0.163^{+ 0.045}_{-0.160} $, is consistent with these two measurements, but can not yet be used to discriminate between them. Interestingly, the posterior distribution of  $z_{c}$, the redshift for which the fractional contribution of the EDE to the total energy content of the Universe is maximal, follows a similar distribution for SPT-3G and Planck, peaking near matter-radiation equality, while we find a slight preference for lower redshift for the ACT data, consistent with previous work~\cite{hill2021}, with a peak closer to decoupling.  

\begin{figure*}[!p]
\includegraphics[width=\textwidth]{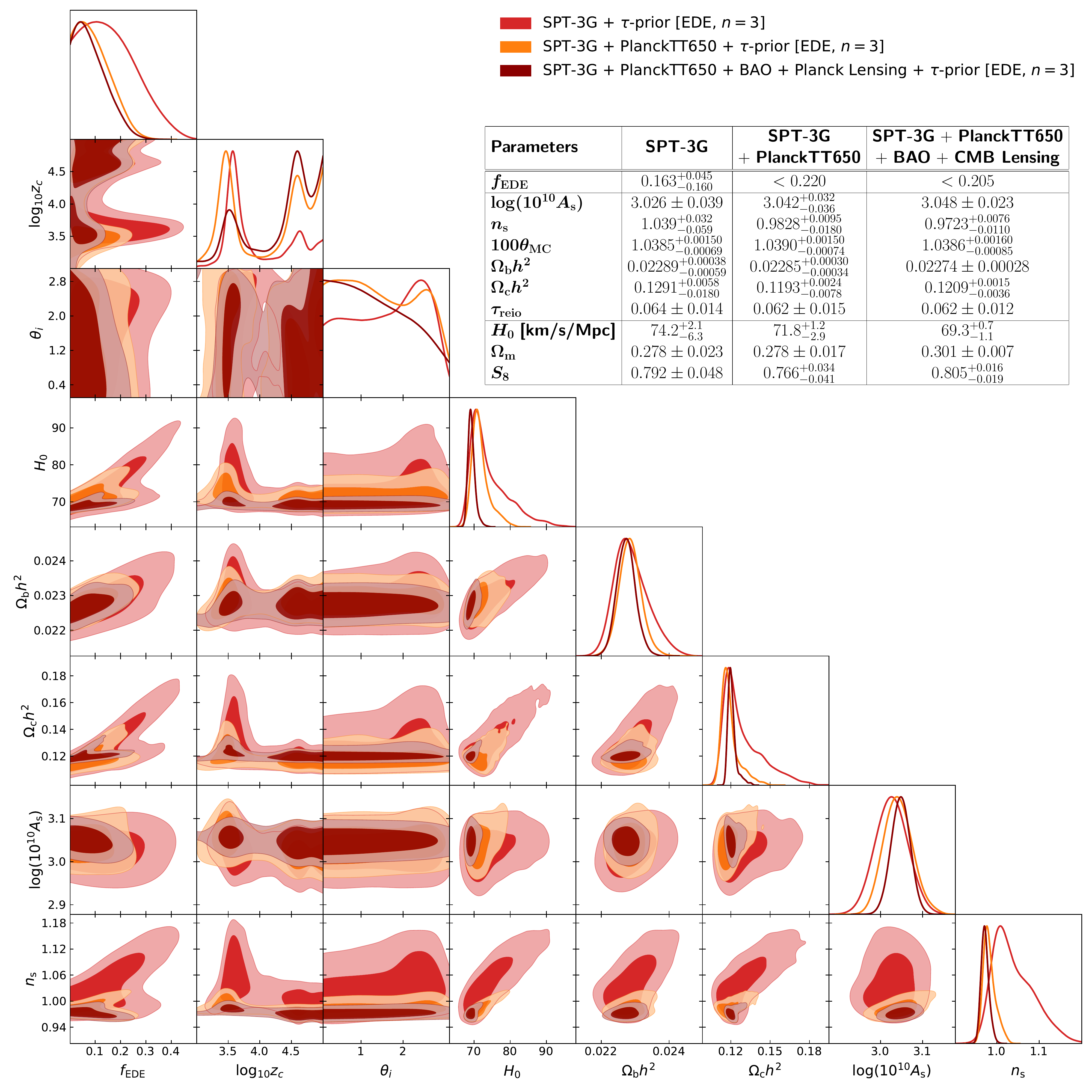}
\caption{Marginalized posterior distributions of EDE ($n=3$) and $\Lambda$CDM parameters derived from SPT-3G (TE+EE) latest data~\cite{dutcher2021measurements} alone (red); in combination with Planck 2018 TT ($\ell<650$) (orange); and combined with Planck 2018 TT ($\ell<650$), Planck CMB lensing and BAO data (6dF~\cite{6dF}, SDSS DR7/DR12~\cite{SDSSDR7,BOSSDR12}) (darker red). We impose the same Gaussian prior on the reionization optical depth $\tau$ for all the analyses. We display the 68\% marginalized constraints and the upper/lower limits (95\% C.L.) for the EDE and $\Lambda$CDM in the table associated with the figure.}\label{fig:spt_only+wmap_matter}
\end{figure*}
\clearpage

We observe with the SPT-3G data the now-familiar $f_{\rm EDE}-H_{0}$ degeneracy that serves to increase the value of $H_{0}$ (and its uncertainties), and reconcile the value of $H_{0}$ inferred from CMB experiments with the ones inferred from local observation of Cepheid-calibrated SNIa. The $\Delta \chi^{2}$ of the EDE scenario with respect to $\Lambda$CDM is of order -4.1 ($\Delta\mathrm{AIC} = 2.1$), which corresponds to a $74.9\%$ C.L. ($1.1\sigma$) preference over $\Lambda$CDM. This does not indicate a significant improvement coming from this scenario given the extra degrees of freedom introduced.

We now consider the following combinations using SPT-3G data:
\begin{itemize}
    \item SPT-3G (TE+EE) + Planck 2018 (TT, $\ell<650$)
    \item SPT-3G (TE+EE) + Planck 2018 (TT, $\ell<650$) + BAO [6dF,SDSS DR7/DR12] + Planck 2018 CMB lensing
\end{itemize}

Posterior distributions and marginalized constraints for these two datasets are shown in Fig.~\ref{fig:spt_only+wmap_matter}. A summary of the best-fit $\chi^2$ values for the studied datasets can be found in Table~\ref{tab:chi2_spt_combin}. Adding the large-scale TT power spectrum from Planck to SPT-3G data tightens the constraints on cosmology (it is particularly noticeable for the scalar index $n_\mathrm{s}$). Large-scale temperature data from Planck also shift the EDE posterior distribution peak towards lower values: we find $f_\mathrm{EDE} < 0.220$ (95\% C.L.) with SPT-3G combined with Planck TT ($\ell<650$), with $\Delta\chi^2 = -4.4$ ($1.2\sigma$) with respect to $\Lambda$CDM. We obtain slightly better constraints on cosmology by adding CMB lensing and BAO measurements, and in particular for the EDE density, we obtain $f_\mathrm{EDE} < 0.205$. We observe a  preference for high $z_c$ values in the $\mathrm{log}_{10}(z_c)$ posterior distribution displayed in Fig.~\ref{fig:spt_only+wmap_matter}, particularly when SPT-3G data are combined with PlanckTT650 and large-scale structure probes. For this dataset, we compute a $\Delta\chi^2$ of $-3.7$ ($1.0\sigma$) with respect to $\Lambda$CDM.\\

\begin{figure}
\includegraphics[width=\linewidth]{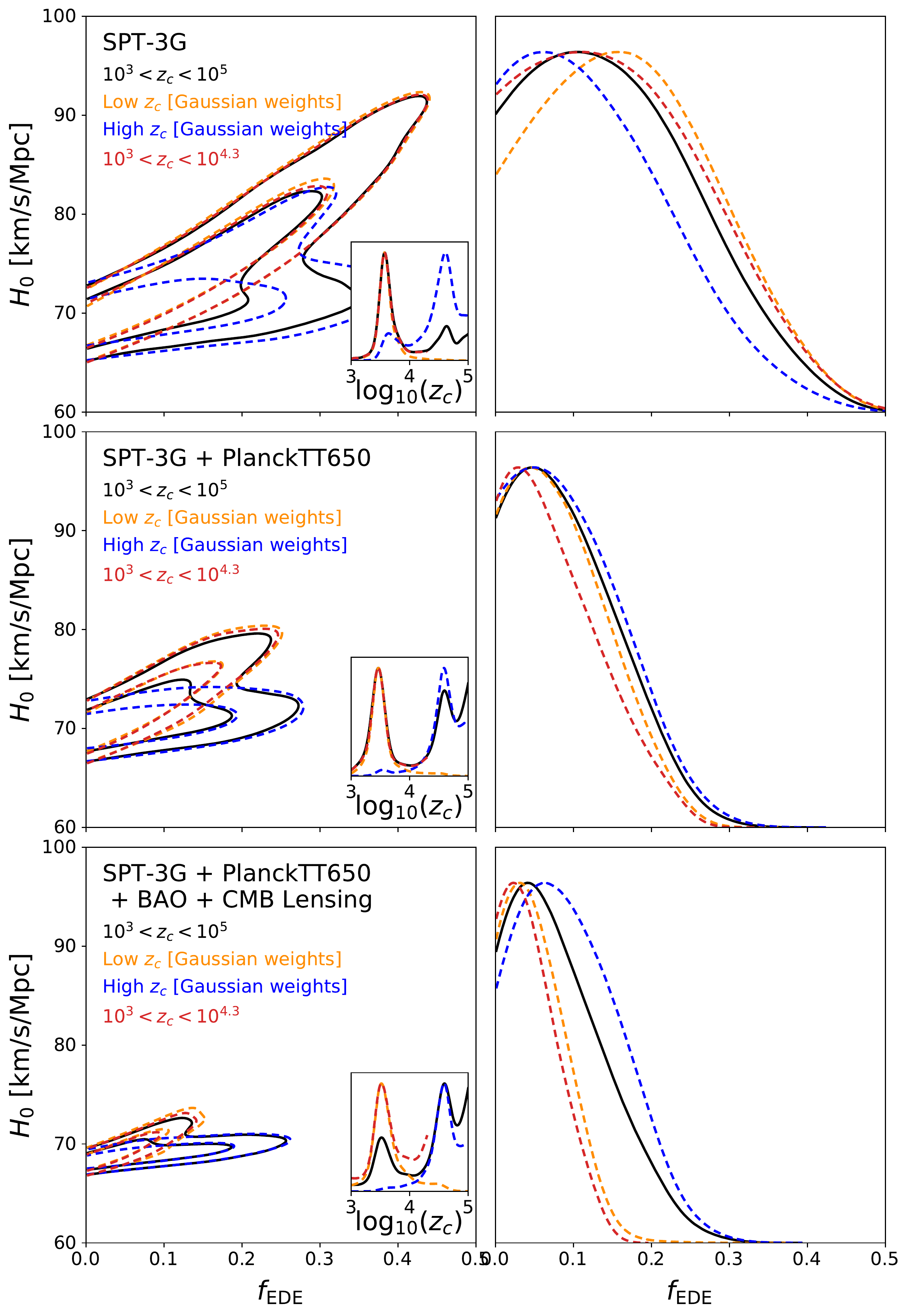}
\caption{\label{fig:fede_h0} 2D contours at 68\% and 95\% C.L. in the reduced parameter space $f_\mathrm{EDE}-H_0$ and $f_\mathrm{EDE}$ 1D posterior distributions derived from the three datasets displayed in Fig.~\ref{fig:spt_only+wmap_matter}: SPT-3G [top panels], SPT-3G + PlanckTT650 [middle panel], and SPT-3G + PlanckTT650 + BAO + Planck Lensing [bottom panel]. We display the 2D posterior distribution using the $z_c$ prior described in Table~\ref{tab:priors} as a black solid line. We apply a re-weighting of the samples according to $z_c$ using Gaussian weights with mean $\mu_\mathrm{low}=3.5$ and standard deviation $\sigma=0.4$ to focus on the samples associated with a small value of $z_c$ (orange). We use Gaussian weights with mean $\mu_\mathrm{high}=4.5$ and standard deviation $\sigma$ to focus on the high $z_c$ samples (blue). We also display the posterior distributions derived from filtered samples using a narrower flat prior on $\mathrm{log}_{10}(z_c)$ (red), which matches the prior range used in~\cite{smith2020,hill2020,Ivanov2020,hill2021}.}
\end{figure}

We study the impact of a high value of $z_c$ on the measurements of other cosmological parameters by re-weighting the samples to favor the higher (lower) $\mathrm{log}_{10}(z_c)$ values using Gaussian weights centered on $\mu_\mathrm{high}=4.5$ ($\mu_\mathrm{low}=3.5$). As shown in Fig.~\ref{fig:fede_h0}, if the EDE field reaches its maximum energy contribution at very early times ($\mathrm{log}_{10}(z_c)\simeq 4.5$), $f_\mathrm{EDE}$ will have a much smaller impact on the sound horizon, and hence the inferred Hubble expansion rate $H_0$, than if it had peaked near the time of recombination. We see this effect on the 2D posterior distributions in the $H_0-f_\mathrm{EDE}$ plane shown in the first column of Fig.~\ref{fig:fede_h0}: the correlation between $H_0$ and $f_\mathrm{EDE}$ tends to disappear when we favor higher $z_c$ values.\\

We observe this behavior when we use CMB lensing and BAO data combined with SPT-3G due to the preference for high $z_c$ values. There are two different degeneracy directions in the 2D posterior distribution of $f_\mathrm{EDE}$ and $H_0$, and the direction with the lowest correlation between these two parameters is favored by this dataset. This leads to low-value of the Hubble parameter $H_0 = 69.3^{+0.7}_{-1.1}$~km/s/Mpc, 3$\sigma$ away from the latest local measurement of $H_0= 73.04 \pm 1.04$~km/s/Mpc from Ref.~\cite{2021arXiv211204510R}. An important takeaway is that the 1D marginalized posterior on $f_{\rm EDE}$ can be quite sensitive to the assumed prior range for $z_c$, and thus assessing evidence for the EDE model based solely on the posterior for this parameter may not always be robust.

\begin{figure*}[!p]
\includegraphics[width=\textwidth]{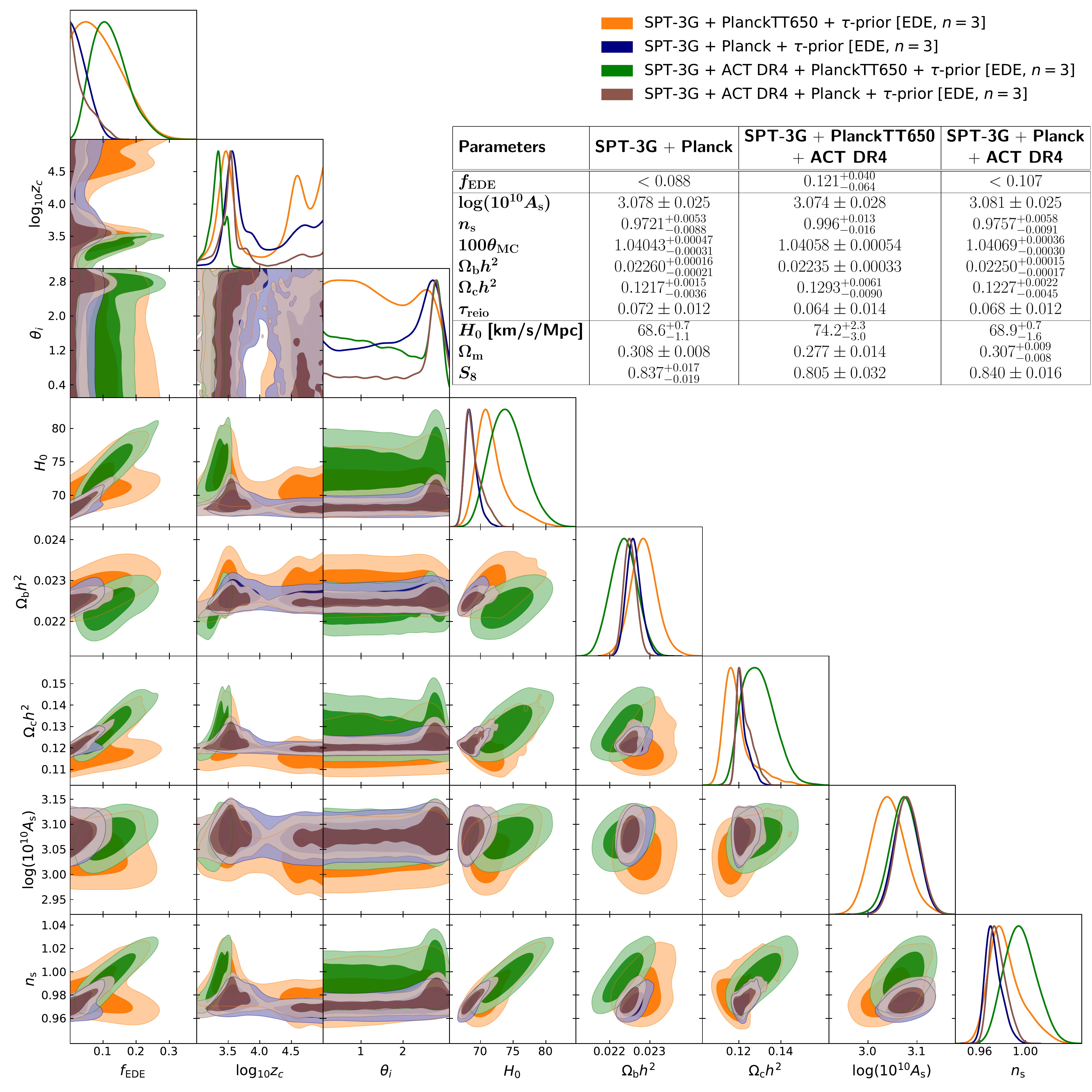}
\caption{Marginalized posterior distributions of EDE ($n=3$) and $\Lambda$CDM parameters derived from SPT-3G (TE+EE) latest data~\cite{dutcher2021measurements} combined with Planck 2018 TT ($\ell<650$) (orange); SPT-3G (TE+EE) combined with the full Planck 2018 dataset (TT+TE+EE) (dark blue); SPT-3G (TE+EE) combined with ACT DR4 (TT+TE+EE)~\cite{Choi2020,Aiola2020} and Planck 2018 TT ($\ell<650$) (green); and SPT-3G (TE+EE) combined with ACT DR4 (TT+TE+EE) and the full Planck 2018 dataset (TT+TE+EE) (brown). We impose the same Gaussian prior on the reionization optical depth $\tau$ for all the analyses. We display the 68\% marginalized constraints and the upper/lower limits (95\% C.L.) for the EDE and $\Lambda$CDM in the table associated with the figure.}\label{fig:spt_act}
\end{figure*}
\clearpage

\begin{table*}[!htbp]
\begin{tabular}{|c|c|c|c|}
\hline
\textbf{Dataset} & & $\mathbf{\Lambda}$\textbf{CDM} & \textbf{EDE} ($\mathbf{n=3}$) \\
\hline\hline
\textbf{SPT-3G} & SPT-3G (TE+EE) & $\mathbf{513.7}$ & $\mathbf{509.6}$\\
$\mathbf{\Delta\chi^2}$ & \dotfill & $$---$$ & $\mathbf{-4.1}$\\
& & & \\

\textbf{SPT-3G + PlanckTT650} & \begin{tabular}{@{}c@{}c@{}} SPT-3G (TE+EE)\\ Planck 2018 high-$\ell$ TT [$\ell<650$]\\ Planck 2018 low-$\ell$ TT\end{tabular} & \begin{tabular}{@{}c@{}c@{}} 516.3 \\ 250.6 \\ 20.8\end{tabular} & \begin{tabular}{@{}c@{}c@{}} 513.1 \\ 250.1 \\ 20.1\end{tabular}\\
& & $\mathbf{787.7}$ & $\mathbf{783.3}$\\
$\mathbf{\Delta\chi^2}$ & \dotfill & $$---$$ & $\mathbf{-4.4}$\\
& & & \\

\begin{tabular}{@{}c@{}} \textbf{SPT-3G + PlanckTT650} \\+ \textbf{BAO + Planck Lensing} \end{tabular} & \begin{tabular}{@{}c@{}c@{}c@{}c@{}} SPT-3G (TE+EE)\\ Planck 2018 high-$\ell$ TT [$\ell<650$]\\ Planck 2018 low-$\ell$ TT\\ BAO [6dF, SDSS DR7/DR12]\\ Planck 2018 Lensing\end{tabular} & \begin{tabular}{@{}c@{}c@{}c@{}c@{}} 516.8 \\ 251.1 \\ 21.7 \\ 7.8 \\ 8.7\end{tabular} & \begin{tabular}{@{}c@{}c@{}c@{}c@{}} 513.8 \\ 249.4 \\ 22.6 \\ 8.0 \\ 8.6\end{tabular}\\
& & $\mathbf{806.1}$ & $\mathbf{802.4}$ \\
$\mathbf{\Delta\chi^2}$ & \dotfill & $$---$$ & $\mathbf{-3.7}$\\
& & & \\

\begin{tabular}{@{}c@{}}\textbf{SPT-3G + PlanckTT650}\\+\textbf{ACT DR4}\end{tabular} & \begin{tabular}{@{}c@{}c@{}c@{}} SPT-3G (TE+EE)\\ Planck 2018 high-$\ell$ TT [$\ell<650$]\\ Planck 2018 low-$\ell$ TT\\ ACT DR4 (TT+TE+EE)\end{tabular} & \begin{tabular}{@{}c@{}c@{}c@{}} 519.2 \\ 251.3 \\ 21.1\\ 295.3\end{tabular} & \begin{tabular}{@{}c@{}c@{}c@{}} 520.7 \\ 249.2 \\ 20.6\\285.0\end{tabular}\\
& & $\mathbf{1086.9}$ & $\mathbf{1075.5}$\\
$\mathbf{\Delta\chi^2}$ & \dotfill & $$---$$ & $\mathbf{-11.4}$\\
& & & \\
\textbf{SPT-3G + Planck} & \begin{tabular}{@{}c@{}c@{}} SPT-3G (TE+EE)\\ Planck 2018 high-$\ell$ (TT+TE+EE)\\ Planck 2018 low-$\ell$ TT \end{tabular} & \begin{tabular}{@{}c@{}c@{}} 520.4\\ 2343.3\\ 23.0\end{tabular} & \begin{tabular}{@{}c@{}c@{}} 519.0\\ 2340.5\\ 21.5\end{tabular}\\
& & $\mathbf{2886.7}$ & $\mathbf{2881.0}$ \\
$\mathbf{\Delta\chi^2}$ & \dotfill & $$---$$ & $\mathbf{-5.7}$\\
& & & \\

\textbf{SPT-3G + Planck + ACT DR4} & \begin{tabular}{@{}c@{}c@{}c@{}} SPT-3G (TE+EE)\\ Planck 2018 high-$\ell$ (TT+TE+EE)\\ Planck 2018 low-$\ell$ TT\\ ACT DR4 (TT+TE+EE) \end{tabular} & \begin{tabular}{@{}c@{}c@{}c@{}} 521.1 \\ 2344.0 \\ 22.5 \\ 244.4\end{tabular} & \begin{tabular}{@{}c@{}c@{}c@{}} 520.2 \\ 2340.9 \\ 21.4 \\ 242.2\end{tabular}\\
& & $\mathbf{3132.0}$ & $\mathbf{3124.7}$ \\
$\mathbf{\Delta\chi^2}$ & \dotfill & $$---$$ & $\mathbf{-7.3}$\\

\hline
\end{tabular}
\caption{\label{tab:chi2_spt_combin} Best-fit $\chi^2$ values computed for $\Lambda$CDM and EDE models and the different dataset combinations. We also display the value of $\Delta\chi^2$ with respect to $\Lambda$CDM in order to quantify the preference for the EDE model.}
\end{table*}

\subsection{Combining with other small-scale CMB data}\label{subsec:sptcombin}

We have shown in Section~\ref{subsec:sptdata} that the constraints on the EDE model from Planck 2018, ACT DR4, and SPT-3G data are consistent, but SPT-3G data are not constraining enough to reach a firm conclusion on the existence of an EDE component in the Universe.  

In this section, we focus on the following combined constraints on the EDE model using the Planck, ACT, and SPT data:

\begin{itemize}
    \item SPT-3G (TE+EE) + Planck 2018 (TT+TE+EE)
    \item SPT-3G (TE+EE) + Planck 2018 (TT, $\ell<650$) + ACT DR4 (TT+TE+EE)
    \item SPT-3G (TE+EE) + Planck 2018 (TT+TE+EE) + ACT DR4 (TT+TE+EE, $\ell_\mathrm{min}^{TT}=1800$)
\end{itemize}

Posterior distributions of EDE and $\Lambda$CDM parameters are shown in Fig.~\ref{fig:spt_act}, and marginalized constraints are displayed in the associated table. To quantify the preference in favor of the EDE model over $\Lambda$CDM, we show the best-fit $\chi^2$ values for these dataset combinations in Table~\ref{tab:chi2_spt_combin}. We do not see any hint in favor of the EDE model in SPT-3G + Planck 2018 data with $f_\mathrm{EDE} < 0.088$ (95\% C.L.). We compute $\Delta\chi^2 = -5.7$ ($\Delta\mathrm{AIC} = 0.3$), corresponding to an insignificant $1.5\sigma$ preference in favor of the EDE model. However, SPT-3G data combined with ACT DR4 and Planck 2018 $TT$ ($\ell<650$) show a hint for non-zero $f_\mathrm{EDE}$ with $f_\mathrm{EDE} = 0.121^{+0.040}_{-0.064}$ (formally 1.9$\sigma$ away from zero). This comes with a higher $\chi^2$ difference between EDE and $\Lambda$CDM: $\Delta\chi^2$ of $-11.4$ ($\Delta\mathrm{AIC} = -5.4$), which corresponds to a $99.0\%$ C.L. ($2.6\sigma$) preference for EDE over the $\Lambda$CDM model. We note that the $\Delta\chi^2$ improvement is mainly driven by ACT DR4 data, for which we compute $\Delta\chi^2 = -10.3$. We also observe that including ACT DR4 data in the analysis provides a strong constraint on $\mathrm{log}_{10}(z_c)$, with a preference for lower values of $z_c$ compared to SPT-3G + PlanckTT650, resulting in a stronger correlation between $f_\mathrm{EDE}$ and $H_0$. Again, the hint for non-zero $f_\mathrm{EDE}$ and the preference in favor of EDE disappears when the full Planck 2018 data are included along with ACT DR4 and SPT-3G data. We find a $95\%$ C.L. upper limit $f_\mathrm{EDE} < 0.107$.

\section{Conclusion}\label{sec:conclusion}

In this work we have used the latest public SPT-3G data set to put constraints on the EDE scenario. We found that the current public data set, representing only a four-month period of observation, is not sensitive enough to reject or favor the presence of a $\sim 10\%$ contribution of EDE to the total energy budget of the Universe near recombination. We expect that forthcoming data release from ACT and SPT, along with the next generation of CMB experiments such as Simons Observatory \cite{Ade2019} and CMB-S4 \cite{Abazajian2016}, should be able to discriminate between such a scenario and $\Lambda$CDM with high statistical significance (see \cite{hill2021} for an initial estimate of the constraining power for upcoming ACT data). \\

\acknowledgments
We thank Gilles Weymann, Stéphane Ili\'c, Evan McDonough, and Michael Toomey for useful comments and discussion. The theoretical power spectra used in this paper were computed using the CAMB Boltzmann solver.  We gratefully acknowledge the IN2P3 Computer Center
(http://cc.in2p3.fr) for providing the computing resources and services needed to perform this work.  JCH thanks the Scientific Computing Core staff at the Flatiron Institute for computational support.  The Flatiron Institute is supported by the Simons Foundation.  JCH acknowledges support from NSF grant AST-2108536.

\begin{appendix}
\section{Early Dark Energy model in \textsc{CAMB}}\label{app:camb_class}
Previous studies of the EDE model using Planck and ACT DR4 data have been performed with modified versions of \textsc{CLASS} \cite{poulin2019,smith2020,hill2020,hill2021,poulin2021}. In the analysis here, we use CMB power spectra computed with the \textsc{CAMB} Boltzmann solver, with an implementation of the EDE model based on Ref.~\cite{smith2020}. 

\begin{figure*}[!p]
\includegraphics[width=\textwidth]{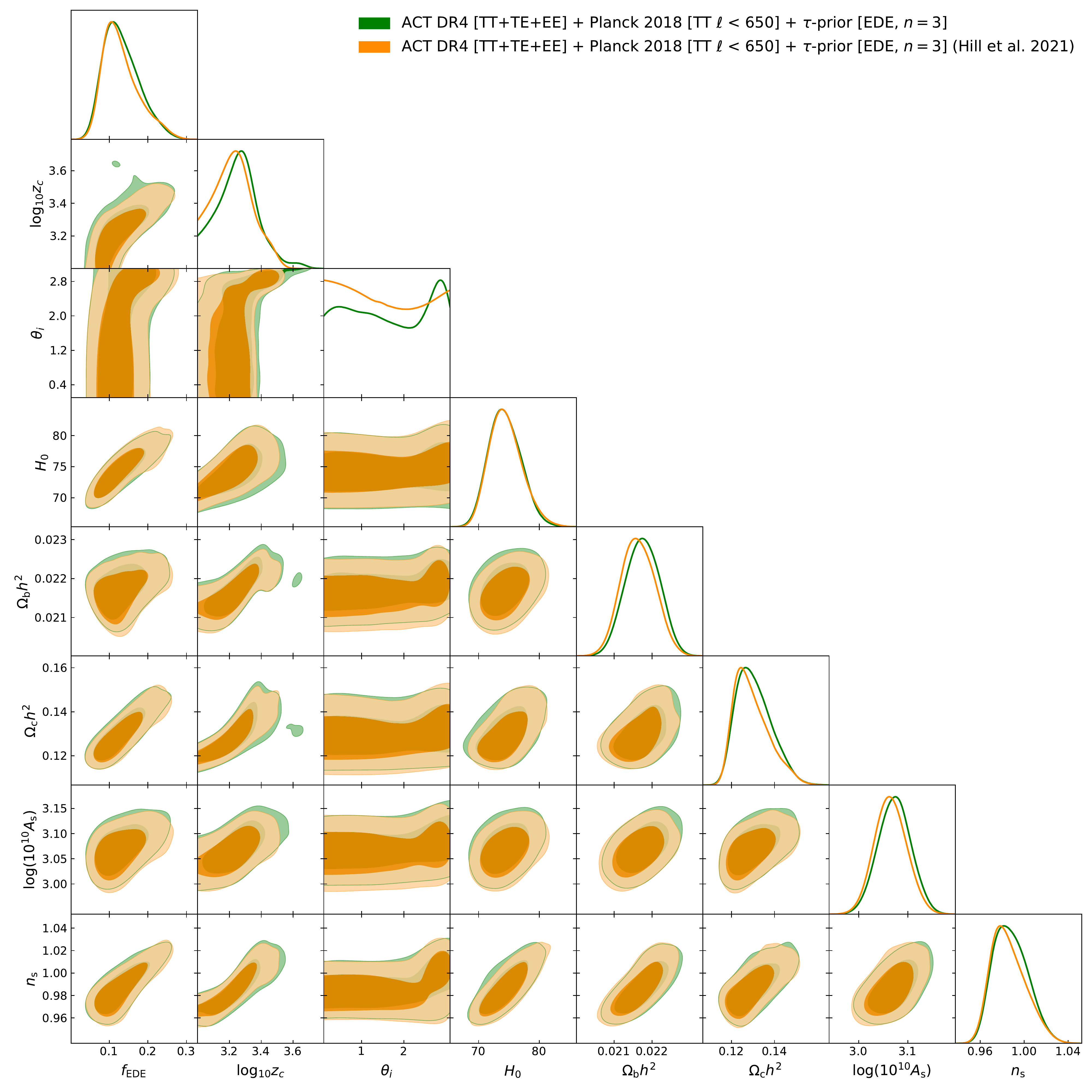}
\caption{Marginalized posterior distributions of EDE ($n=3$) and $\Lambda$CDM parameters derived from ACT DR4 (TT+TE+EE) and Planck 2018 TT ($\ell < 650)$ using the \textsc{CAMB} Boltzmann solver (green) compared to the results obtained using \textsc{CLASS\_EDE}~\cite{hill2020} in Ref.~\cite{hill2021} (orange). In both cases, we impose the same Gaussian prior on the reionization optical depth $\tau$.
}\label{fig:act+p18tt650_hill_vs_camb}
\end{figure*}

We reproduce the results of Ref.~\cite{hill2021} using the EDE implementation in \textsc{CAMB}.  We use ACT DR4 (TT+TE+EE) in combination with the lowest multipoles of the Planck 2018 TT power spectrum ($\ell<650$).  The posterior distributions of EDE and $\Lambda$CDM parameters are displayed in Fig.~\ref{fig:act+p18tt650_hill_vs_camb}.

We find that the posterior distributions of both EDE and $\Lambda$CDM parameters are in excellent agreement between the two implementations of the EDE scalar field model, and we recover a $f_\mathrm{EDE}$ posterior distribution that peaks $2.4\sigma$ away from zero in ACT DR4 + PlanckTT650 data with $f_\mathrm{EDE}=0.132^{+0.034}_{-0.056}$. We observe an improvement in the $\chi^2$ with respect to $\Lambda$CDM with $\Delta\chi^2 = -16.1$, corresponding to a $99.9\%$ C.L. ($3.3\sigma$) preference in favor of the EDE model over $\Lambda$CDM mainly driven by the ACT DR4 data. This has already been observed in Ref.~\cite{hill2021}, with $\Delta\chi^2 = -15.4$ ($99.8\%$ C.L. or $3.2\sigma$ using the same datasets.

\end{appendix}
\bibliography{main}


\end{document}